\documentclass[usenatbib,letterpaper]{mn2e}
\usepackage[totalwidth=480pt,totalheight=680pt,layoutvoffset=0.5cm]{geometry}
\usepackage{times,amssymb,hyperref,subfigure,epsfig,aas_macros}

\begin{document}

\title[Nature or nurture of coplanar Tatooines]{Nature or nurture of coplanar Tatooines: the
  aligned circumbinary Kuiper belt analogue around HD~131511}

\author[Grant M. Kennedy] {Grant M. Kennedy\thanks{Email:
    \href{mailto:gkennedy@ast.cam.ac.uk}{gkennedy@ast.cam.ac.uk}} \\
  Institute of Astronomy, University of Cambridge, Madingley Road, Cambridge CB3
  0HA, UK \\
}
\maketitle

\begin{abstract}
  A key discovery of the \emph{Kepler} mission is of the circumbinary planets known as
  ``Tatooines'', which appear to be well aligned with their host stars' orbits. Whether
  this alignment is due to initially coplanar circumbinary planet-forming discs
  (i.e. nature), or subsequent alignment of initially misaligned discs by warping the
  inner disc or torquing the binary (i.e. nurture), is not known. Tests of which scenario
  dominates may be possible by observing circumbinary Kuiper belt analogues (``debris
  discs''), which trace the plane of the primordial disc. Here, the 140 au diameter
  circumbinary debris disc around HD~131511 is shown to be aligned to within 10$^\circ$
  of the plane of the near edge-on 0.2 au binary orbit. The stellar equator is also
  consistent with being in this plane. If the primordial disc was massive enough to pull
  the binary into alignment, this outcome should be common and distinguishing nature
  versus nurture will be difficult. However, if only the inner disc becomes aligned with
  the binary, the HD~131511 system was never significantly misaligned. Given an initial
  misalignment, the $\sim$Gyr main-sequence lifetime of the star allows secular
  perturbations to align the debris disc out to 100 au at the cost of an increased scale
  height. The observed debris disc scale height limits any misalignment to less than
  25$^\circ$. With only a handful known, many more such systems need to be characterised
  to help test whether the alignment of circumbinary planets is nature or nurture.
\end{abstract}

\begin{keywords}
  stars: binaries --- circumstellar matter --- planets and satellites: formation ---
  planet-disc interactions --- stars: individual: HD~131511 (HIP~72848)
\end{keywords}

\section{Introduction}\label{s:intro}

Planet formation appears to be a near ubiquitous process, being successful in single and
multiple star systems
\citep{1995Natur.378..355M,2003ApJ...599.1383H,2011Sci...333.1602D}, and across a wide
range of stellar host masses
\citep{1998A&A...338L..67D,2002ApJ...576..478F,2006Natur.439..437B}. It is therefore
surprising that the discoveries of the first circumbinary planets were only made recently
\citep{2011Sci...333.1602D}, when the first evidence for successful circumbinary
planetesimal formation was revealed 30 years ago by the discovery of a Kuiper belt
analogue around the eclipsing A-type binary $\alpha$ CrB
\citep{1985PASP...97..885A}. This extended wait does not reflect a paucity of
circumbinary planets however \citep{2014MNRAS.444.1873A}, merely a strong bias towards
discovery of the lower hanging fruit that are planets around stars in single and wide
multiple systems. Indeed, aside from their perhaps inevitable discovery with
\emph{Kepler} \citep[e.g.][]{2011Sci...333.1602D}, few surveys are specifically targeting
nearby close binary stars with the goal of disentangling their spectra to push down to
mass limits that are competitive with the state of the art around single stars
\citep[e.g.][]{2005ApJ...626..431K,2009ApJ...704..513K}.

The known circumbinary planets are well aligned with their binary host orbits, but this
alignment is a heavy bias towards their discovery in eclipsing binary systems. An
analysis of circumbinary planets finds that the occurrence rate is similar compared to
single stars, but only if those planets are typically coplanar with the binary
\citep{2014MNRAS.444.1873A}. If the planets have a wider range of inclinations, the
occurrence rate goes up, and hence is higher than the single star rate. Given theoretical
work that finds circumbinary planet formation is if anything harder than around single
stars \citep[e.g.][]{2004ApJ...609.1065M,2007MNRAS.380.1119S,2012ApJ...754L..16P}, an
enhanced circumbinary planet occurrence rate would be a surprise, and a high degree of
coplanarity seems more likely.

Such coplanarity is also expected, at least in the inner regions where these planets
form, as a result of the torques exerted on a young circumbinary protoplanetary disc by
the binary \citep[and vice versa,][]{2013ApJ...764..106F,2013MNRAS.433.2142F}. Whether
coplanarity is expected farther out in the disc is less clear; it may be that alignment
is simply a natural outcome of binary star formation, but interaction with the
surrounding environment, for example the late infall of gas onto the protoplanetary disc
may lead to misalignment. Whether this alignment can be corrected is uncertain, depending
on poorly constrained parameters such as disc mass and viscosity
\citep[e.g.][]{2000MNRAS.317..773B,2013ApJ...764..106F,2013MNRAS.433.2142F}. The
alignment of inner ($\lesssim$10 au) disc regions is potentially probed empirically by
the known circumbinary planets. However, the state of more distant regions, beyond where
the known close-in planets probably form, relies on the detection of small-body
populations; the gas-poor Kuiper belt analogues (``debris discs'') seen around
main-sequence binaries that are thought to trace the original plane of the protoplanetary
disc.

Here, star-binary-disc alignment in the spectroscopic binary system HD~131511 is
considered. The debris disc in this system has been studied previously, focussing on
either the disc structure \citep{2014arXiv1408.5649M} or whether the disc has an
inclination similar to that inferred for the primary star \citep{2014MNRAS.438L..31G}. A
more detailed analysis of the disc geometry is used here to show that all system
components are plausibly aligned. The ability of binary perturbations to bring an
initially misaligned disc towards coplanarity during the primordial and debris disc
phases, and the implications for any initial misalignment, are then considered.

\section{The HD~131511 system}

\subsection{The Stars}

HD~131511 (HIP~72848) is a single-lined spectroscopic binary with a K0V primary
\citep{2003AJ....126.2048G}, whose orbit was first derived using radial velocities
\citep{1981JRASC..75...56K,1983PASP...95...79B,2013AJ....145...41K}. The near to edge-on
geometry of the orbit was subsequently derived using Hipparcos astrometry
\citep{2005A&A...442..365J}, and the derived elements, along with the age
\citep{2008ApJ...687.1264M} and masses estimated from the $B-V$ colour are reproduced in
Table \ref{tab:bin}. Using the observed rotation period \citep[$10.39 \pm 0.03$
days,][]{1995AJ....110.2926H}, $v\sin i$ \citep[$4 \pm 1$ km
s$^{-1}$,][]{1990ApJS...72..191S}, and stellar radius derived from fitting stellar
atmosphere models, \citet{2014MNRAS.438L..31G} found that the primary star is inclined by
at least 70$^\circ$, and though the position angle of the projected spin axis is unknown
this calculation shows that the equator of the primary is consistent with being aligned
with the binary orbit.

\begin{table}
  \caption{HD~131511 binary properties; orbit
    \citep{2002ApJS..141..503N,2005A&A...442..365J}, age \citep{2008ApJ...687.1264M} and
    masses \citep{2005A&A...442..365J}. The ascending node $\Omega$ is measured East of
    North.}\label{tab:bin} 
  \begin{tabular}{llll}
    \hline
    Parameter & Symbol (unit) & Value & $1\sigma$ \\
    \hline
    Semi-major axis & a (mas) & 16.54 & 0.18 \\
    Semi-major axis & a (au) & 0.19 & 0.03 \\
    Eccentricity & e & 0.51 & 0.001 \\
    Inclination & i ($^\circ$) & 93.4 & 4.2 \\
    Ascending node & $\Omega$ ($^\circ$) & 248.3 & 3.6 \\
    Longitude of pericenter & $\omega$ ($^\circ$) & 219 & 0.1 \\
    Orbital Period & P (days) & 125.396 & 0.001 \\
    Distance & d (pc) & 11.51 & 0.06 \\
    Mass of A & $M_{\rm A} (M_\odot)$ & 0.79 & -  \\
    Mass of B & $M_{\rm B} (M_\odot)$ & 0.45 & -  \\
    Age & Gyr & 1 & 0.3 \\
    \hline
\end{tabular}  
\end{table}

\subsection{The Debris Disc}\label{s:obs}

Most debris discs are detected via infrared (IR) excesses above the level expected from
the stellar photosphere. HD~131511 was reported to have an infrared excess at 24 $\mu$m
by \citet{2010ApJ...710L..26K}. However, this was due to an incorrect prediction for the
photospheric flux density, presumably from the use of saturated 2MASS data (they found
133 mJy, where the true value is closer to 200 mJy). This error was corrected by
\citet{2013ApJ...768...25G}, who found no 24 $\mu$m excess, but reported a 3.2$\sigma$
significant 100 $\mu$m excess. A similar level of excess at 70 $\mu$m is also seen,
meaning that the excess is robust.

The disc is in fact resolved with the \emph{Herschel} Photodetector Array Camera and
Spectrometer
\citep[PACS,][]{2010A&A...518L...1P,2010A&A...518L...2P},\footnote{\emph{Herschel} is an
  ESA space observatory with science instruments provided by European-led Principal
  Investigator consortia and with important participation from NASA.} and the properties
have been reported in \citet{2014MNRAS.438L..31G} and
\citet{2014arXiv1408.5649M}. \citet{2014MNRAS.438L..31G} derived just the inclination of
the disc in order to compare it with the stellar inclination, while
\citet{2014arXiv1408.5649M} modelled the radial structure, finding that the images lack a
signal to noise ratio sufficient to constrain more than the radial disc location.

\begin{figure*}
  \begin{center}
    \hspace{-0.25cm} \includegraphics[width=0.5\textwidth]{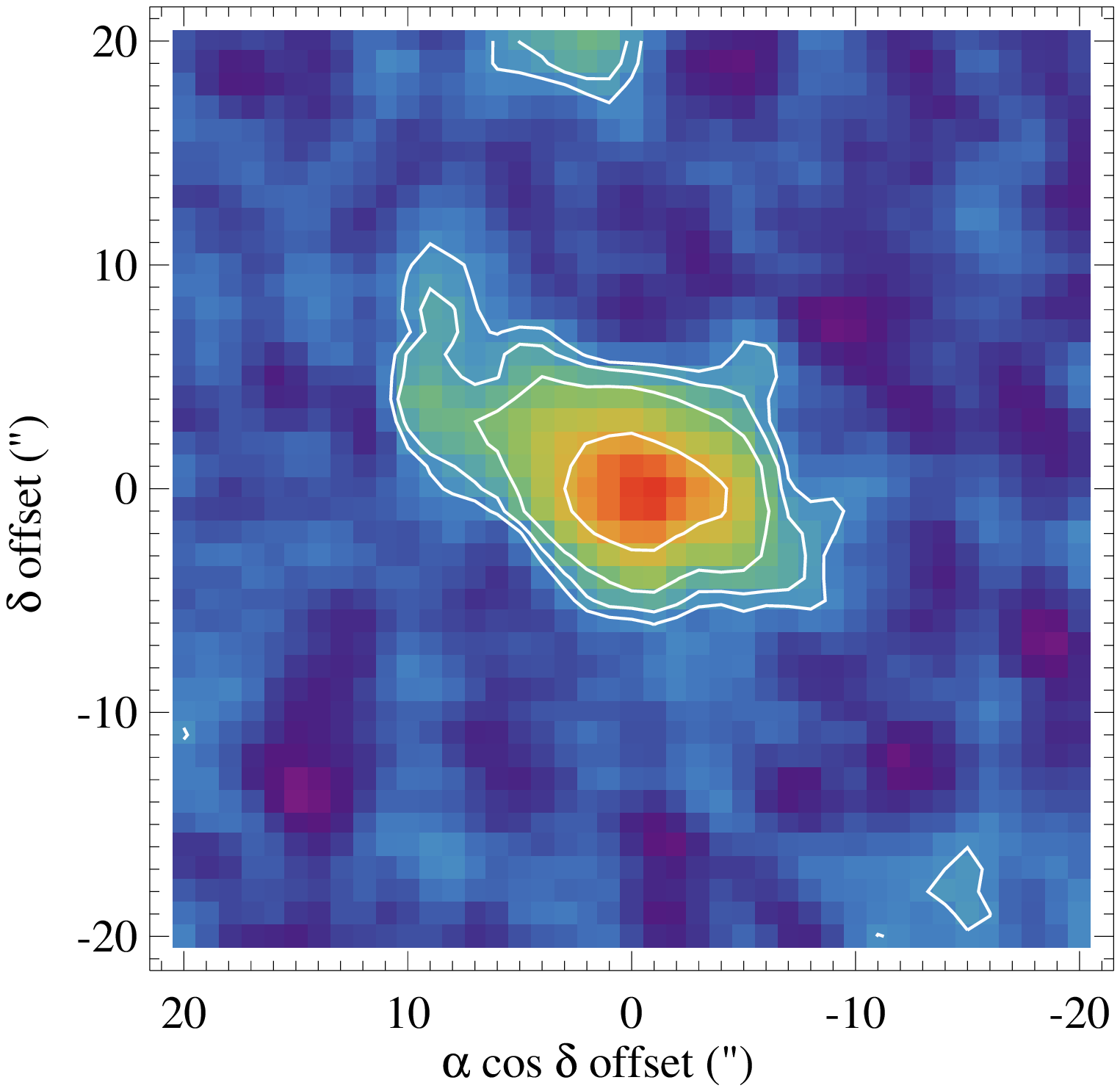}
    \hspace{-2cm} \includegraphics[width=0.5\textwidth]{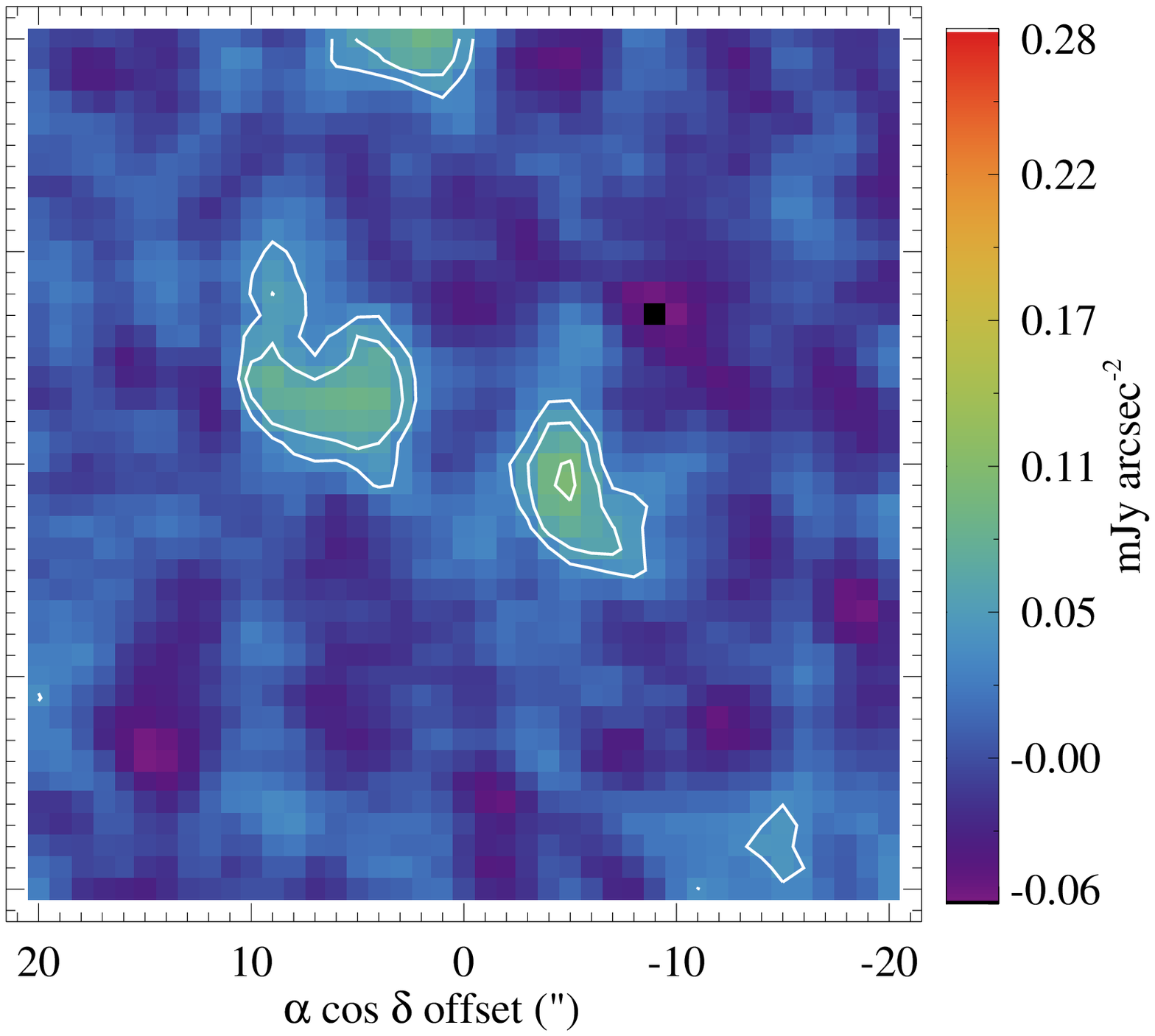}
    \caption{\emph{Herschel} 100$\mu$m images of HD~131511, North is up and East is
      left. The left panel shows the raw image with contours at 2, 3, 5, and 10 times the
      pixel RMS of $1.9 \times 10^{-2}$ mJy arcsec$^{-2}$. The right panel shows the same
      image and contours after a peak-scaled point source has been subtracted, leaving
      residuals at the disc ansae as a clear sign of a resolved disc that is near to
      edge-on.}\label{fig:ims}
  \end{center}
\end{figure*}

The disc structure is best constrained at 100$\mu$m where the disc/star contrast is
highest, and original and star-subtracted PACS images at this wavelength are shown in
Fig. \ref{fig:ims}. These data are the same as used by both \citet{2014MNRAS.438L..31G}
and \citet{2014arXiv1408.5649M} so the reader is referred to those papers for details
regarding data reduction. \citet{2014arXiv1408.5649M} concluded that the disc is
consistent with being a radially narrow ring with a radius of 60-70 au. Both studies
concluded that the disc is highly inclined ($>$70$^\circ$) and a position angle of
66$^\circ$ East of North was derived in the latter study.

\section{System geometry and alignment}

\begin{figure}
  \begin{center}
    \hspace{-0.5cm} \includegraphics[width=0.5\textwidth]{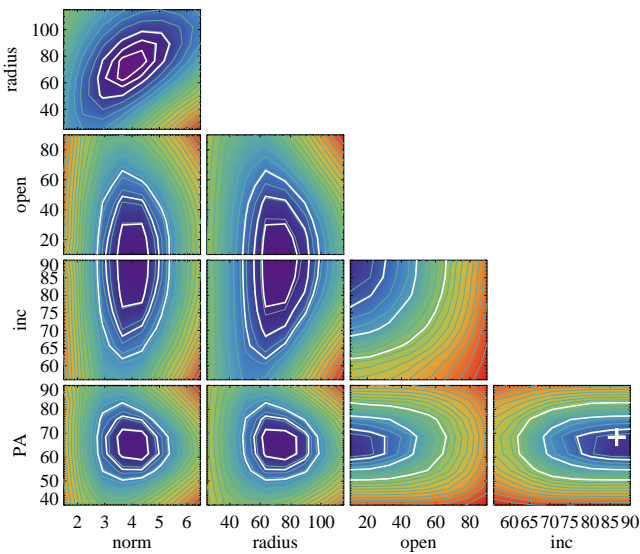}
    \caption{Significance contours for varying disc normalization (arbitrary units),
      radius (in au), disc opening angle, inclination, and PA (all in degrees). Each
      panel shows contours for two parameters when marginalised over the other
      three. White contours are 1, 2, and 3$\sigma$. The white cross marks the
      orientation of the HD~131511 binary orbit.}\label{fig:cont}
  \end{center}
\end{figure}

The derived position angle and inclination of the disc is consistent with the ascending
node and inclination of the binary orbit, and the inclination of the primary star's spin
axis. There is of course an ambiguity in the ascending node of the debris disc, because
there is currently no way to infer the side of the star on which the dust is coming
towards us, so it could be that the planetesimal disc in fact orbits in the opposite
sense to the binary. Using the same dust ring model that was used in
\citet{2014MNRAS.438L..31G} and \citet{2014arXiv1408.5649M}, this level of consistency is
quantified in more detail in Fig. \ref{fig:cont}, which shows a series of 2-dimensional
marginalizations over a 5-dimensional grid search for the best fitting disc parameters
that reproduce the image in Fig. \ref{fig:ims}. The parameters varied were the brightness
and radius of a 10 au-wide dust ring, the vertical opening angle of the dust belt, and
the position angle and inclination of the belt. At each location in this space, a
high-resolution disc model was created and convolved with the PACS 100 $\mu$m beam (an
observation of $\gamma$ Dra), and the $\chi^2$ goodness-of-fit metric then computed from
the difference between the model and the data.

The lack of a strong disc signal means that a simple narrow ring is a good fit to the
data \citep{2014arXiv1408.5649M}, which has a radius of about 70 au, but could be as
small as 50 au or as large as 100 au. The contours in Fig. \ref{fig:cont} show the 1, 2,
and 3 $\sigma$ confidence regions computed from the $\Delta$$\chi^2$ compared to the best
fit disc parameters. The white cross in the lower right panel shows the binary orbit
position angle and inclination. The disc and binary are therefore consistent with being
aligned, but at the 2 $\sigma$ level the disc could be up to about 20$^\circ$ less
inclined than the binary, and have a position angle about 10$^\circ$ different. The disc
opening angle is less than about 50$^\circ$. Though the position angle of the star is
unknown, the possible inclination discrepancy between the binary orbit and stellar
equator is around 20$^\circ$.

Thus, this analysis shows for the first time that the circumbinary debris disc around
HD~131511 is consistent with being aligned with the binary orbital plane. By quantifying
the uncertainties in important disc parameters, the main conclusions of this modelling
are that the disc opening angle could be as large as 50$^\circ$, and that any undetected
misalignment could be as large as 20$^\circ$.

\section{Discussion}

Having noted that HD~131511 is consistent with having the equator of the primary star,
the 0.2 au binary orbit, and the 70 au radius debris disc in the same plane, the origin
and dynamics of the system, and other similar systems, are considered.

\subsection{Protoplanetary disc alignment}

Circumbinary protoplanetary disc formation may be a messy process, and the possibility of
late gas infall means that binary-disc coplanarity is not a guaranteed initial condition
\citep[e.g.][]{2013ApJ...764..106F}. The probable coplanarity as inferred from the
\emph{Kepler} planets suggests that at a minimum the inner regions of the planet-forming
discs can become aligned with the binary, and thus the system loses the signature of any
initial disc misalignment. The timescale with which alignment occurs, and the degree and
extent of alignment, is uncertain. \citet{2013ApJ...764..106F} explored a low-viscosity
case, finding that the disc is not strongly warped, and that the binary is pulled into
alignment with the binary on a timescale comparable with the lifetimes of primordial
discs, albeit with considerable uncertainty due to very strong dependence on (for
example) the disc scale height and inner edge location. With larger viscosity,
\citet{2013MNRAS.433.2142F} find that large warps are possible, and that the disc inner
edge is not generally aligned with the binary (though they restrict their analysis to
disc masses low enough that the binary orbit is unaffected).

Thus, the degree to which circumbinary protoplanetary discs are typically aligned with
binary orbits, and the timescale for any alignment, is uncertain. The estimates of
\citet{2013ApJ...764..106F} suggest that an initially misaligned disc may not exhibit a
large warp, and that the binary can become aligned with the primordial disc via
interaction at the inner edge before it is dispersed. If this case is typical then all
discs around close binaries could become aligned before they disperse, and perhaps before
planets form. However, the disc may not have enough mass to pull the binary into
alignment, and may also have a significant warp. In this case the disc becomes aligned
with the binary relatively slowly, though the timescales are
uncertain. \citet{2000MNRAS.317..773B} estimate that alignment may occur on the same
timescale as the secular nodal precession induced in the disc by the binary (discussed
below), though depending on the origin of damping within the disc \citep[e.g. if the
parametric instability does not operate][]{2000MNRAS.318.1005G} could also be a lot
longer.

\begin{figure}
  \begin{center}
    \hspace{-0.5cm} \includegraphics[width=0.5\textwidth]{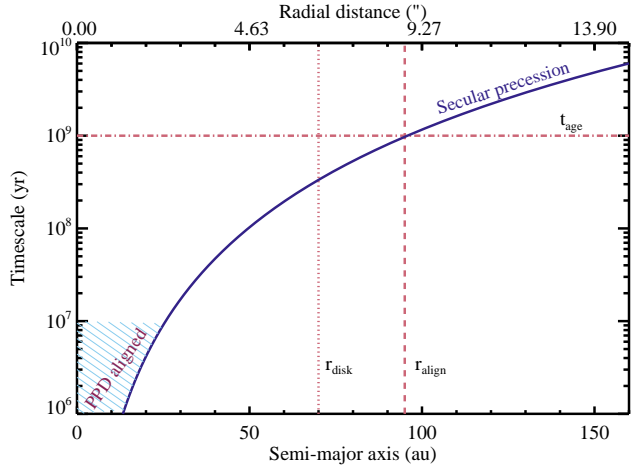}
    \caption{Protoplanetary disc alignment and secular precession times for circumbinary
      orbits as a function of semi-major axis. A circumbinary protoplanetary disc can be
      aligned within a few tens of au during the $\sim$10$^7$ year disc
      lifetime. Planetesimals can on average become aligned (i.e. have executed one full
      cycle of secular precession) within $\sim$1 Gyr if they reside within 95
      au. Therefore, the debris disc may have become aligned due to secular perturbations
      on the main-sequence.}\label{fig:tsec}
  \end{center}
\end{figure}

For the specific example of HD~131511, the secular precession time is shown in
Fig. \ref{fig:tsec}, calculated according to \citet{2010MNRAS.401.1189F} using the binary
parameters from Table \ref{tab:bin}. An estimate for where the primordial disc can become
aligned with the binary is shown by the hatched region, assuming that the alignment
timescale is the same as the precession time, and allowing for disc lifetimes up to
$10^7$ years \citep[e.g.][]{1995Natur.373..494Z,2006ApJ...651.1177P}. With this estimate,
disc regions beyond a few tens of au do not become aligned within reasonable
protoplanetary disc lifetimes. Compared to the primordial disc lifetimes, the much longer
main-sequence lifetimes of stars means that the debris disc that forms in, and then
emerges from the protoplanetary disc can be strongly affected by secular perturbations to
much larger radial distances.

\subsection{Debris disc alignment}

In the absence of the dissipation present in a gaseous disc secular, perturbations from
the binary cause disc particle inclinations and lines of nodes to oscillate rather than
damp, with a timescale that depends on semi-major axis. The eccentricities and pericenter
arguments also vary, but are less relevant because the eccentricities imposed depend
inversely on the separation ratio between the particles and the binary
\citep[e.g.][]{2004ApJ...609.1065M}, and are hence very small at $\sim$70 au around
HD~131511 ($\lesssim$10$^{-3}$).

The effect of the perturbations on the disc particles can be viewed in two ways. If the
reference plane is pictured as that of the original disc plane (so the binary is
inclined), then the inclinations of the particles oscillate about the binary plane as
their lines of nodes circulate (libration is possible for large initial misalignments and
high binary eccentricities). This is the picture usually applied in planetary systems,
for example the warp in the $\beta$ Pictoris disc \citep[e.g.][]{1997MNRAS.292..896M}. If
the reference plane is taken to be that of the binary, the particle orbits are initially
inclined, and precess about the binary angular momentum vector. Only if the binary is
eccentric do their inclinations also change as they precess
\citep[e.g.][]{2010MNRAS.401.1189F,2012MNRAS.421.2264K}. Thus, given enough time any disc
will appear to become aligned at the cost of an increased scale height, with an opening
angle equal to twice the initial misalignment.

\begin{figure}
  \begin{center}
    \hspace{-0.5cm} \includegraphics[width=0.5\textwidth]{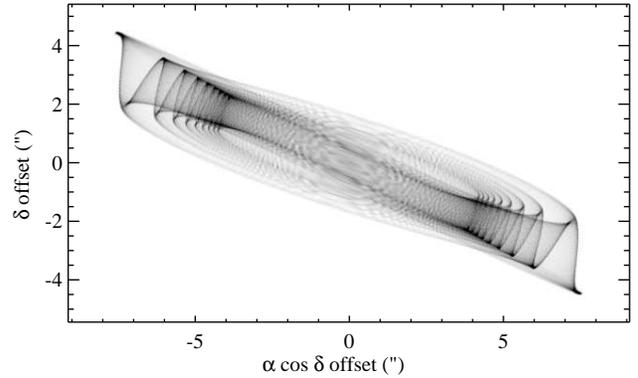}
    \caption{Simulation of mm-wave structure for an initially flat 50-100 au disc with
      particles on circular orbits around the HD~131511 binary after 1Gyr. The initial
      misalignment was 10$^\circ$ with a line of nodes with respect to the binary
      pericenter of 0$^\circ$. A blackbody temperature law and 1mm wavelength was assumed
      in creating the image.}\label{fig:secim}
  \end{center}
\end{figure}

The secular precession time, the time taken for a particle to undergo one full
oscillation in inclination (and a complete circulation of the line of nodes), for
particles orbiting HD~131511 at a range of distances is shown in Fig. \ref{fig:tsec}. The
estimated disc radius of 70 au is shown, and at the stellar age of around 1 Gyr particles
out to about 95 au can have their inclinations vary significantly. Thus, the alignment of
the disc with the binary orbit is not necessarily primordial, but given that the disc is
seen to have an opening angle smaller than about 50$^\circ$, any initial misalignment
must have been smaller than about 25$^\circ$.

Fig. \ref{fig:secim} shows a synthetic mm-wave image of a 50-100 au circumbinary disc
around HD~131511 as it would appear if the initial misalignment was 10$^\circ$. The wavy
structure arises as the particles' orbits' nodes precess, becoming less distinct closer
to the star where the precession is fastest and the most precession cycles have taken
place. Observed with a resolution greater (i.e. worse) than a few seconds of arc, such a
disc would simply appear to be aligned with the binary. At higher resolution
($\sim$1\arcsec) the vertical structure would be resolvable and hence the initial
misalignment could be inferred or constrained from the scale height. At yet higher
resolution the wavy radial structure might be seen, though whether it actually exists or
is smoothed out would depend on various complicating factors that are not included in
this model, such as the (uncertain) eccentricity of the disc particles and whether
radiation or stellar wind forces strongly affect mm-size dust.

\subsection{Other circumbinary debris discs}

Aligned circumbinary discs were previously found around HD~98800BaBb
\citep{2010ApJ...710..462A}, $\alpha$ CrB (HD~139006), and $\beta$ Tri
\citep[HD~13161,][]{2012MNRAS.426.2115K}, while a misaligned disc has been found around
99 Her \citep{2012MNRAS.421.2264K}. In the cases of $\alpha$ CrB and $\beta$ Tri the disc
as resolved with \emph{Herschel} is sufficiently well separated from the binary that
perturbations do not affect it within the stellar lifetime and that the alignment is
primordial. The same cannot be said for the HD~98800 hierarchical quadruple system; the
disc that orbits the BaBb pair is relatively compact, probably due to truncation by the
AaAb pair \citep{2010ApJ...710..462A}. Perturbations from the BaBb pair probably protect
the disc from strong perturbations from the AaAb pair
\citep{2008MNRAS.390.1377V,2009MNRAS.394.1721V}, though greater reddening seen towards
the disc hosting pair may be a sign of disc warping and that this protection is not
absolute \citep{2007ApJ...670.1240A}. In any case, dynamics in the HD~98800 system will
be more complex if enough gas is present, and several studies suggest that the system is
in fact in the late stages of dispersing a gaseous planet-forming disc rather than a
``true'' gas-poor debris disc \citep{2007ApJ...664.1176F,2012ApJ...744..121Y}.

Adding HD~131511 to this sample, the debris discs seen around three (four including
HD~98800) close binaries are seen to be aligned, while the single wider case of 99~Her is
strongly misaligned. With such a small sample the trend can simply be noted, as can the
goal of building the numbers in order to make quantitative statements about the origins
of circumbinary alignment. With nearly all resolved debris discs residing in systems
within a few hundred parsecs, GAIA is the most promising in this regard, both for
discovering and characterising binary systems, and also for discovering circumbinary
planets \citep{2014arXiv1410.4096S}. The best case scenario would see discoveries of
systems where coplanarity tests of planet and disc orbits could be made across a wide
range of radial distances.

\section{Conclusions}\label{s:disc}

The measurable components of the HD~131511 system are aligned. The binary orbit is well
known, and the debris disc geometry is consistent with being in the same plane. While the
position angle of the stellar spin axis is unknown, the inclination is also consistent
with the equator being aligned with the binary orbit. The timescale for alignment during
the primordial gaseous phase may be too long compared to the gas disc lifetime beyond a
few tens of au, so if the debris disc traces the plane of the gas disc any initial
misalignment was limited to less than 25$^\circ$. The sample of systems where such tests
have been made remains small, and many more are needed to make a strong test of whether
the alignment of circumbinary planet orbits is nature or nurture.

\section{Acknowledgements}

GMK is supported by the European Union through ERC grant number 279973. Thanks to Stefano
Facchini for helpful discussions on primordial circumbinary disc warping and to the
referee for a swift and helpful review.


\end{document}